\documentstyle[preprint,aps]{revtex}
\tightenlines
\begin{document}
\draft
\title{Negative hopping magnetoresistance of two-dimensional electron
gas in a smooth random potential}
\author{M. E. Raikh,$^{(1)}$ L. I. Glazman$^{(2)}$}
\address{
$^{(1)}$Physics Department, University of Utah, Salt Lake City, Utah
84112\\
$^{(2)}$Theoretical Physics Institute and Department of Physics, University of
Minnesota, Minneapolis,  Minnesota 55455}
\maketitle
\begin{abstract}
We show that the tunnel coupling between semiclassical states localized in
different minima of a smooth random potential increases when magnetic field is
applied. This increase originates from the difference in gauge factors which
electron wave functions belonging to different electron ``lakes'' acquire in
the
presence of the field. We illustrate the increase of coupling by a model
calculation of tunneling through a saddle point separating two adjacent lakes.
In
the common case, when the barrier between two lakes is much narrower than their
size, the characteristic magnetic field is determined by the area of the lakes,
and
thus may be quite small. The effect of the field on coupling constants leads to
a
negative magnetoresistance in low-temperature conduction.
\end{abstract}
Deep in the insulating regime, the low-temperature conductivity of a
two-dimensional electron gas is dominated  by phonon-assisted electron hops
between
localized states. The conventional picture\cite{shklovskii} of hopping implies
that
localized states are formed by individual impurities. The lower is the
temperature,
the larger is the separation of the localized states between which a typical
hop
takes place. In this picture, the magnetic field affects hopping transport in
two
ways: by shrinking the wave functions of the localized states, and by changing
the
contributions of different tunneling paths to the amplitude of a hop. The
latter
means that in the course of tunneling between initial and final states an
electron
can ``visit'' a sequence of virtual states localized on neighboring impurities,
so
that the amplitude of a hop represents a sum of partial amplitudes. In a
magnetic
field, each path acquires a phase factor. These factors suppress the
destructive
interference of different paths, thus causing the decrease\cite{Nguen,Imry} of
the
resistance in a weak magnetic field. It was shown in\cite{Nguen} that this
decrease
comes from specific hops for which tunneling amplitudes are close to zero in
the
absence of a field. Although the portion of these hops is small, the rise of
conductivity with magnetic field is related to the increase of the tunneling
probabilities for these particular hops, since they are most sensitive to the
field.

A different realization of the insulating regime emerges if the random
potential is
smooth. In the latter case, the electron gas breaks up into separate lakes,
each
lake accommodating many electrons. Within a certain temperature range the
electron
transport is provided by tunneling of electrons through saddle points of
barriers
separating the adjacent lakes. This picture is different from the standard
picture
of the variable-range-hopping\cite{shklovskii}, so that the straightforward
application of the theory of magnetoresistance\cite{Nguen,Imry} is impossible.
In
the present paper, we study magnetoresistance of electron gas in this regime.

It is important to note that the states active in transport (i.e., with
energies
close to the Fermi level) correspond to high-number levels of size quantization
in
the minima of the random potential. As a result, the amplitude of a hop is
determined not only by the transmission coefficient at the saddle point, but
also
depends significantly on the overlap between the wave functions on the both
sides
of the barrier. This overlap is normally small because of the oscillatory
behavior
of the wave functions. If the applied magnetic field is weak enough, its prime
role
is to affect this overlap. We argue that the overlap {\sl increases} with
magnetic
field, i. e., the coupling between the states, localized in adjacent minima of
the
potential, becomes stronger. In other words, unlike the
conventional\cite{Nguen,Imry} model, magnetoresistance of each elementary link
appears to be negative thus leading to the net negative magnetoresistance of
the
system.

Consider two minima separated by a barrier. In the vicinity of a saddle
point the barrier potential has the form:
\begin{equation}
U(x,y)=U_0-\frac{m\omega_x^2 x^2}{2}+\frac{m\omega_y^2 y^2}{2},
\label{barrier}
\end{equation}
see Fig. 1. We define the coupling constant between the states $p$ and $k$
in the neighboring minima as a matrix element $t_{pk}$ in the tunneling
Hamiltonian:
\begin{equation}
H=\sum_{p}E_p^la_p^\dagger a_p + \sum_{k}E_k^rb_k^\dagger b_k
 + \sum_{pk}(t_{pk}a_p^\dagger b_k + t_{pk}^*b_k^\dagger a_p),
\label{tunneling}
\end{equation}
where $E_p^l$, $E_k^r$ are the energy levels in the two minima, and
$a_p^\dagger$, $b_k^\dagger$ are the corresponding creation operators. The
expression for $t_{pk}$ in terms of the wave functions $\psi_p^l(x,y)$ and
$\psi_k^r(x,y)$ can be obtained by generalizing to the two-dimensional case
of the procedure employed in the derivation of the tight-binding
approximation\cite{statphys}:
\begin{equation}
t_{pk}=\frac{\hbar^2}{2m}\int\int_{-\infty}^\infty
dy_1dy_2G_E(y_1,y_2)
\frac{\partial}{\partial x}\psi_p^l\left(-\frac{d}{2},y_1\right)
\left[\frac{\partial}{\partial
x}\psi_k^l\left(\frac{d}{2},y_2\right)\right]^\star.
\label{t}
\end{equation}
Here $m$ is the electron mass, and $G_E(y_1,y_2)$ is the Green function
describing the propagation of an electron under the barrier between the
points $(-d/2,y_1)$ and $(d/2,y_2)$, see Fig. 1. In the derivation of
(\ref{t}) we have neglected the curvature of the lines of turning points,
$U(x,y)=E$, on the both sides of the barrier. The form of the barrier
(\ref{barrier})
enables one to separate the variables in the region $|x|<d/2$.  As a result the
Green function can be expressed in terms of eigenfunctions $\varphi_n(y)$ of
the
harmonic oscillator with frequency $\omega_y$,
\begin{equation}
G_E\left(y_1, y_2\right) = \sum_n
\frac{\varphi_n(y_1) \varphi_n(y_2)}
{u_n'(d/2)v_n'(-d/2) - v_n'(d/2)u_n'(-d/2)},
\label{ge}
\end{equation}
where $u_n(x)$ and $v_n(x)$  of the one-dimensional Schroedinger equation with
the barrier potential $U_0-m\omega_x^2x^2/2$ and energy
$E-(n+1/2)\hbar\omega_y$. If the transmission coefficient of the barrier is
small, one can use the quasiclassical expressions for the functions in
(\ref{ge}),
$u_n,v_n\propto \exp\{\pm \int_0^x dx^{\prime} \kappa_n (x^{\prime})\}/\sqrt
{\kappa_n}$ with
\begin{equation}
\kappa_n = \frac{\sqrt{2m}}{\hbar} \left[U_0 -
\frac{m\omega^2_xx^2}{2} - E+\hbar\omega_y (n+\frac{1}{2})\right]^{1/2}.
\label {kappa}
\end{equation}

The precise form of the potential confining the electrons in each lake is a
complicated functional of the distribution of positive charges outside the
plane of
$2D$ electron system. If the size of a lake exceeds the effective Bohr radius,
$a_B=\hbar^2\varepsilon/me^2$, screening by the electrons of the lake
renormalizes
considerably the bare confining potential (here $\varepsilon$ is the dielectric
constant). However, for a {\sl two-dimensional} electron gas, the screened
potential has still the spatial scale of variations equal to the lake size
rather
than being equal to a small screening radius\cite{scr}. Therefore, we use the
simplest  parabolic form for the confinement potentials in order to carry out
the
calculations to the end. We assume
\begin{eqnarray}
&&V_l(x, y) = \frac{m\Omega^2_{lx}(x+D_1+d/2)^2}{2} +
\frac{m\Omega^2_{ly}y^2}{2}, \label{vl} \\
&&V_r(x, y) = \frac{m\Omega^2_{rx}(x-D_2-d/2)^2}{2} +
\frac{m\Omega^2_{ry}y^2}{2} \nonumber
\end{eqnarray}
for the potentials forming the left and right lakes respectively. The
corresponding
equipotential lines are elliptic with excentricity depending on the
corresponding
ratios $\Omega_{(l,r)x}/\Omega_{(l,r)y}$.  Then the eigenfunctions in each lake
are
the products of the harmonic oscillator wave functions in $x$ and
$y$-directions.  Denote with $Y_{l}$ and $Y_{r}$ the turning points of the wave
functions in the $y$-direction for the states $p$ and $k$ respectively (see
Figure~1). If $Y_{l}$ and $Y_{r}$ exceed significantly the effective size of
the
tunneling region, the $y$-dependence of $\psi^l_p(-d/2, y_1)$ and
$\psi^r_k(d/2,y_2)$ takes  the simple form:
\begin{eqnarray}
&&\psi^l_p\left(-\frac{d}{2}, y_1\right)\propto
\sin\left(\frac{m}{\hbar}\Omega_{ly}Y_ly_1+\chi_l\right), \label{psi} \\
&&\psi^r_k\left(\frac{d}{2}, y_2\right)\propto
\sin\left(\frac{m}{\hbar}\Omega_{ry}Y_ry_2+\chi_r\right), \nonumber
\end{eqnarray}
where the phases of $\chi_l$ and $\chi_r$ are $0$ or $\pi/2$ depending on the
parity of the numbers of the harmonic oscillator wave function. Obviously, such
a
simple form applies when both numbers are high.

As an example, we discuss the coupling between two even (in the $y$-direction)
states
$\chi_l = \chi_r = \pi/2$.  Then Eq. (\ref{t}) reduces to the evaluation of the
following double integral:
\begin{eqnarray}
t_{pk} &\propto & \int dy_1 \int dy_2
\cos\left(\frac{m}{\hbar}\Omega_{ly}Y_ly_1\right)
\cos\left(\frac{m}{\hbar}\Omega_{ry}Y_ry_2\right)\nonumber\\
&\times & \sum_{n} \varphi_n(y_1)\varphi_n(y_2)
\exp\left(-\int_{-d/2}^{d/2}dx'\kappa_n(x')\right).
\label{tpk}
\end{eqnarray}
Note that the integrals over $y_1$ and $y_2$ can be considered as the Fourier
transforms of the functions $\varphi_n$.  Since for a harmonic oscillator the
Fourier
transform of the eigenfunction has the same functional form as $\varphi_n$, we
have
\begin{equation}
t_{pk}\propto
\sum_{n}\varphi_{2n} \left(Y_l\frac{\Omega_{ly}}{\omega_y}\right)
\varphi_{2n}\left(Y_r\frac{\Omega_{ry}}{\omega_y}\right)
\exp\left( -\int_{-d/2}^{d/2}dx^\prime\kappa_{2n}(x^\prime)\right).
\label{tpk2}
\end{equation}
The summation over $n$ can be carried out if we assume the term
$(n+\frac{1}{2})\hbar\omega_y$ in Eq.~(\ref{kappa}) to be a small correction
and expand
$\kappa_n$ with respect to this term.  Then we have
\begin{equation}
\int_{-d/2}^{d/2} dx'\kappa_n(x')\approx \frac{\pi (U_0-E)}{\hbar \omega_x}
+\pi (n+ \frac{1}{2})\frac{\omega_y}{\omega_x}.
\label{int}
\end{equation}
Now we can make use of the addition rule\cite{Beitmen} for the functions
$\varphi_n$ to
obtain:
\begin{equation}
t_{pk}\propto
\exp \left[ - \frac{\pi(U_0-E)}{\hbar\omega_x} -
\frac{m(Y^2_{l}\Omega^2_{ly} + Y^2_{r}\Omega^2_{ry})}
{2\hbar\omega_y \tanh (\pi\omega_y/\omega_x)}\right]
\cosh \left(\frac{mY_{l}Y_{r}\Omega_{ly}\Omega_{ry}}
{\hbar\omega_y \sinh (\pi\omega_y/\omega_x)}\right).
\label{tkp}
\end{equation}
The first term in the exponent describes the transmission amplitude for the
barrier at
the central crossection $y=0$.  The other terms describe the reduction of
$t_{kp}$ due
to the oscillations of the wave functions $\psi_p$ and $\psi_k$.  The expansion
of
$\kappa_n$ (and, hence, Eq(\ref{tkp})) is justified if the main contribution to
$\ln t_{kp}$ comes from the tunneling exponent.

The magnetic field $B$ affects the coupling matrix element $t_{kp}$ by
modifying the
wave functions $\psi_p$ and $\psi_k$ in the lakes and by changing the Green
function
Eq.~(\ref{ge}) for the motion under the barrier.  However, if we choose the
guage in
the  form
\begin{equation}
A_x = 0,\;  A_y = Bx,
\label{ax}
\end{equation}
then the Green function will not be affected (if the width
of the tunneling region is small compared to the lake's size). Therefore, we
have
to find the variation with magnetic field of the wave functions $\psi_p$,
$\psi_k$
only.  Let us start from the left lake centered at $x=-d/2 - D_1$.  It is
convenient first to find the function $\tilde{\psi}_{p}(x,y)$ in the gauge $A_x
= 0$,
$A_y = B(x+ d/2 + D_1)$ and then to make a transformation to the initial
guage Eq.(\ref{ax}).  As a result of this transformation, the wavefunction
$\psi_{p}$
acquires a gauge phase factor:
\begin{equation}
\psi_p (x, y) = \tilde{\psi}_p (x, y)\exp\left[
\frac{2\pi iB}{\Phi_0}\left(\frac{d}{2}+D_1\right)y\right],
\label{psi1}
\end{equation}
$\Phi_0$ being the flux quantum.  A similar consideration is applicable to the
wave
functions in the right lake, and those acquire an additional factor $\exp[-2\pi
iBy(d/2 + D_2)/\Phi_0]$.

It is convenient to write the Schroedinger equation for the function
$\tilde{\psi}_{p}$ using the variable $x_1 = x + D_1 + \frac{d}{2}$ instead of
$x$.  Then we have
\begin{equation}
-\frac{\hbar^2}{2m}\left(\frac{\partial^2 \tilde{\psi}_p}{\partial x_{1}^{2}} +
\frac{\partial^2 \tilde{\psi}_p}{\partial y^2}\right)
- i\hbar\omega_c x_1 \frac{\partial \tilde{\psi}_p}{\partial y}
+\frac{m}{2} \left[({\Omega}^2_{lx}+\omega_c^2)x^{2}_{1}
+\Omega^2_{ly}y^2\right] \tilde{\psi}_p = E\tilde{\psi}_p,
\label{hbar}
\end{equation}
where $\omega_c = eB/cm$ is the cyclotron frequency.  We search for the
quasiclassical solution of Eq.~(\ref{hbar}) in the form $\exp[iS(x, y)]$, where
$S$ is the action.  For weak enough magnetic fields, $\omega_c\ll\Omega_{lx},
\Omega_{ly}$ we can present $S(x,y)$ as $S_{0}+S_{1}$, where $S_{0}$ is the
action for motion in the potential $m(\Omega^2_{lx}x^2 + \Omega^2_{ly}y^2)/2$
defined by the equations:
\begin{equation}
\frac{\partial S_0}{\partial x_1} = \frac{m \Omega_{lx}}{\hbar}
\sqrt{D_1^2 - x_1^2},\; \frac{\partial S_0}{\partial y} =
\frac{m\Omega_{ly}}{\hbar} \sqrt{Y_l^2 - y^2}.
\label{partialso}
\end{equation}
Assuming $S_1\ll S_0$ we get the following equation for the correction
$S_{1}(x,y)$:
\begin{equation}
\frac{\hbar^2}{m} \left(\frac{\partial S_0}{\partial x_1}
\frac{\partial S_1}{\partial x_1} + \frac{\partial S_0}{\partial y}
\frac{\partial S_1}{\partial y}\right)
= - m\omega_c\Omega_{ly}x_1 \sqrt{Y_l^2 - y^2}.
\label{hbar2}
\end{equation}
It is easy to check that the solution of this equation has the form
\begin{equation}
S_1(x_1, y)= \frac{-m\omega_c\Omega_{ly}}{\hbar (\Omega^2_{ly}-\Omega_{lx}^2)}
\left[\Omega_{ly}x_1y + \Omega_{lx} \sqrt{D_1^2 - x_1^2}
\sqrt{Y_l^2 - y^2}\right].
\label{s}
\end{equation}
Eq.~(\ref{s}) gives the phase which a state $p$ acquires in the magnetic field.
 To
calculate the coupling coefficient with the help of Eqs.~(\ref{t})
and(\ref{psi1}), we
need the value of $S_1$ at $x_1=D_1$. It is convenient to present the phase
factors
[originating from (\ref{psi1}) and (\ref{s})] for the wave functions
$\psi_p(-d/2,
y_1)$ and $\psi_k(d/2, y_2)$ as
$\exp (2\pi i y_1 L_1/\Phi_0)$ and $\exp (-2\pi i y_2 L_2/\Phi_0)$, where
\begin{equation}
L_{1,2} = \frac{d}{2} + \frac{\Omega_{(l,r)x}^2}
{\Omega_{(l,r)x}^2 - \Omega_{(l,r)y}^2}D_{1,2}.
\label{calL}
\end{equation}
After introducing these factors into Eq~(\ref{t}), the calculation of the
coupling
coefficient is quite similar to the case of zero magnetic field, and yields:
\begin{equation}
t_{kp}(B) = t_{kp}(0) \exp \left(-\frac{B^2}{2B^2_0}\right) \cosh\frac{B}{B_1}.
\label{tkp3}
\end{equation}
It is convenient to express the parameters $B_0$ and $B_1$ in terms of
$L_1$, $L_2$, and two amplitudes of zero-point motion,
$\lambda_s=\sqrt{\hbar/m\omega_y}$ and
$\lambda_{l,r}=\sqrt{\hbar/m\Omega_{(l,r)x}}$, that characterize respectively
the
saddle point of the barrier and the lakes,
\begin{eqnarray}
B_0 & = & \frac{\Phi_0}{\sqrt2\pi\lambda_s\sqrt{L_1^2+L_2^2}}
\left[\frac{1}{\tanh (\pi\omega_y/\omega_x)}+
\frac{2L_1L_2}{(L_1^2+L_2^2)\sinh (\pi\omega_y/\omega_x)}\right]^{-1/2},
\label{B_0}\\
B_1 & = & \frac{\Phi_0}{2\pi L_1L_2}
\left[\left(\frac{\lambda_s}{\lambda_l}\right)^2\frac{Y_l}{L_1}
\left(1+\frac{L_1}{L_2} \cosh \frac{\pi\omega_y}{\omega_x}\right)-
\left(\frac{\lambda_r}{\lambda_s}\right)^2\frac{Y_r}{L_2}
\left(1+\frac{L_2}{L_1}\cosh\frac{\pi\omega_y}{\omega_x}\right)\right]^{-1}
\sinh\frac{\pi\omega_y}{\omega_x}.
\nonumber
\end{eqnarray}
Correspondingly the magnetoresistance between two lakes behaves as
\begin{equation}
\frac{R(B)}{R(0)} = \left(\frac{t_{kp}(B)}{t_{kp}(0)}\right)^{-2} =
\exp\left(\frac{B^2}{B_0^2}\right) \frac{1}{\cosh^2 B/B_1}.
\label{R}
\end{equation}
The magnetoresistance at small $B$ is quadratic,
\begin{equation}
\frac{\delta
R(B)}{R(0)}=\left(\frac{1}{B_0^2}-\frac{1}{B_1^2}\right)B^2,\nonumber
\end{equation}
and it is negative if $B_{0}>B_{1}$. To estimate the ratio
$B_0/B_1$, we notice that typically all the dimensionless factors in
(\ref{B_0}) are
of the order of unity, and therefore $B_0/B_1\sim L_{1,2}/\lambda_s$. The
latter ratio
can be estimated as the number of occupied levels in each of the lakes, and is
large
in the semiclassical regime. Hence, in the small field region the resistance
decreases
as $\delta R(B)/R(0)\simeq -B^2/B_1^2$, the characteristic value $B_{1}$ being
of the
order of $\Phi_0/D_1D_2$. The resistance falls off exponentially with $B$ in
the
region $B_c \gg B \gg B_1$, where the crossover field is $B_c=B_0^2/B_1$. After
reaching a minimum at $B\approx B_c$, the resistance rises sharply at higher
fields. At $B \gg B_c$ the exponential increase of the resistance persists. The
origin
of magnetoresistance in this limit can be understood in terms of the edge
states  formation. Indeed, it can be easily seen that under the condition
$B \gg B_c$ the cyclotron radius becomes smaller than the size of the lake.
Two edge states on the opposite sides of the barrier correspond to the orbits
skipping in {\em opposite } directions. Then the mismatch of the wave vectors
characterizing these orbits becomes only bigger as the magnetic field
increases. Note however that Eq~(\ref{R}) is not valid at $B \gg B_c$ since at
$B=B_c$
 the cyclotron frequency $\omega_c$ becomes of the order of $\Omega_x$

The physical meaning of the negative magnetoresistance is that the phase
factors
acquired by the wavefunctions in magnetic field compensate effectively the
difference in their wave numbers in the $y$-direction and, therefore,
 lead to the
increase of the coupling. This is clearly seen from the Eq.~(\ref{t}). Indeed,
the
Green function $G_E(y_1,y_2)$ connects most effectively points with $y_1=y_2$,
and
$G_E(y,y)$ is a slow function of its argument. On the other hand, functions
$\psi_k^l(d/2,y)$ and $\psi_p^r(d/2,y)$ rapidly oscillate with {\sl different}
periods. We have shown explicitly on a simple model, that a weak magnetic field
reduces the ``distance'' between the Fourier harmonics belonging to these two
wavefunctions thus increasing the overlap between $\partial\psi_l/\partial x$
and
$\partial\psi_r/\partial x$ at the barrier. A similar phenomenon for the
``vertical''
tunneling between parallel two-dimensional electron systems was considered
recently by Lian Zheng and MacDonald\cite{MacDonald}.

The general case of lakes of an arbitrary shape does not allow one to
separe the variables, which may lead to
``quantum chaos''\cite{Gutzwiller,Stone,Harold}. It might be argued\cite{Berry}
then, that the states $\psi_l$ and $\psi_r$ are well represented at each
point by a random superposition of plane waves travelling in all directions.
This superposition, along with other harmonics, would contain the plane wave
striking the inter-lake barrier at right angle. Due to the latter harmonic, the
coupling matrix elements $t_{k,p}$ would loose the exponential sensitivity to
the magnetic field. However the results of the recent studies (see the book by
Gutzwiller\cite{Gutzwiller}, and  Heller's chapter in\cite{LesHouches} and
references therein) suggest that even in the case of chaotic motion, a
substantial portion of quantum eigenstates with high numbers are associated
with
the periodic classical trajectories in the lake. Namely, these states can be
conceived as the result of the Bohr-Sommerfeld quantization of  plane waves
propagating along these trajectories. The numbers of states of the both types
described above are of the same order\cite{Baranger}. It means that strong
negative magnetoresistance and exponential suppression of conductivity
persists for random shapes of the electronic lakes, because of the existence of
``scarred''\cite{Gutzwiller,LesHouches} eigenstates.

The magnetoresistance (\ref{R}) has no temperature dependence. This is because
while deriving our main result (\ref{B_0}), (\ref{R}) we assumed the motion of
electrons within each lake to be completely coherent. At a finite temperature,
the
phase-breaking length, $L_\phi$, may become smaller than $D_{1,2}$. This will
eliminate effectively, the parts of the phase factors (\ref{calL})
 acquired due to the electron motion within each of the lakes, i.e. the terms
in
$L_{1,2}$ proportional to $D_{1,2}$. Therefore at $L_\phi\lesssim D_{1,2}$
$L_{1,2}$ in (\ref{calL}) should be replaced by the value of $d/2$. As a
result, in
the range of temperatures where $L_\phi\sim D_{1,2}$, the magnetoresistance
must
exhibit a strong temperature dependence, and become weaker at higher
temperatures.
The sign of the magnetoresistance remains negative, because for a semiclassical
electron motion $d$ exceeds $\lambda_{l,r,s}$, and the condition
$B_0\gg B_1$ remains valid, even if $L_\phi\lesssim D_{1,2}$.

In conclusion we have demonstrated that the hopping conductivity of
two-dimensional electron gas subjected to a smooth random potential
should increase drastically with increasing magnetic field. The larger are the
electron lakes (the closer is the Fermi level to the classical percolation
threshold), the sharper is  the raise of the conductivity with  magnetic field.
The giant negative hopping magnetoresistamce in weak magnetic fields that may
be attributed to the mechanism we considered here, was recently reported for
GaAs gated heterostructures\cite{Ji,W}.
However, the observed\cite{Ji,W} transition to the metallic (Quantum Hall)
state
at higher fields can not be accounted for within this simple model.

Discussions with  I. L. Aleiner, B. L. Altshuler and H. U. Baranger are
gratefully acknowledged. The work at the University of Minnesota was
supported by NSF Grant DMR-9423244.

\begin{figure}
\caption{Two electronic lakes separated by a saddle point ({\it a}), and
the schematic potential profile in the crossection y=0 ({\it b})}
\label{fig1}
\end{figure}
\end{document}